\documentclass[aps, prb, reprint]{revtex4-1}
\usepackage{graphicx}
\usepackage{epstopdf}
\usepackage{amsmath, amssymb}

\begin{document}

\title{Broadband chiral metamaterials with large optical activity}

\author{Kirsty Hannam$^1$}
\email{kirsty.hannam@anu.edu.au}
\author{David A. Powell$^1$}
\author{Ilya V. Shadrivov$^1$}
\author{Yuri S. Kivshar$^{1,2}$}

\affiliation{$^1$Nonlinear Physics Center, Research School of Physics and Engineering,
Australian National University, Canberra, ACT 0200, Australia}

\affiliation{$^2$National Research University of Information Technology, Mechanics and Optics,
St Petersburg 197101, Russia}

\begin{abstract}
We study theoretically and experimentally a novel type of metamaterial with hybrid elements composed of twisted pairs of cross-shaped meta-atoms and their complements.  We reveal that such two-layer metasurfaces demonstrate large, dispersionless optical activity at the transmission resonance accompanied by very low ellipticity. We develop a retrieval procedure to determine the effective material parameters for this structure, which has lower-order symmetry ($\mathrm {C}_4$) than other commonly studied chiral structures. We verify our new theoretical approach by reproducing numerical and experimental scattering parameters.
\end{abstract}

\maketitle

\section{Introduction}

Chiral structures with optical activity and circular dichroism have been instrumental for many applications including biological and chemical sensing~\cite{Rogachevaetal2006}. In particular, chiral metamaterials can have optical activity several magnitudes of order larger than the effects found in nature. A chiral metamaterial formed by twisting planar structures, such as a pair of crosses or split-ring resonators, can result in large optical activity or giant gyrotropy at a range of frequencies~\cite{Rogachevaetal2006,Deckeretal2009, Weietal2011b}. The resonant modes of these twisted structures will be dominated by either an electric or magnetic dipole response leading to the impedance being mismatched to free space. Also, optical activity is highly dispersive over the transmission band, and it should be accompanied by ellipticity due to the Kramers-Kronig relations~\cite{Weietal2011b,Deckeretal2010,Hendryetal2012,Zhaoetal2012,Lietal2012}. This is undesirable for many polarization-based applications requiring linearly polarized light.

The Babinet principle states that an infinitely thin, perfectly conducting complementary structure illuminated by a complementary incident field generates a field equivalent to the field excited in the original structure but with the electric and magnetic fields exchanged~\cite{electromag,Falconeetal2004,Bitzeretal2011,Lietal2011}. Intuitively, by coupling an element together with its complement these electric and magnetic responses become coupled, matching the impedance over the transmission peak, which should overcome the previously stated short-comings in rotated structures.

The approach based on combining a meta-atom with its complement has previously been used to study non-chiral effects, such as dual-band ultra-slow modes~\cite{NavarroCiaetal2010} and a broad bandpass filter at THz frequencies~\cite{Chiangetal2011}. The coupling mechanisms of this approach have also been studied at optical frequencies, and circular dichroism observed~\cite{Hentscheletal2013}.

Previously, we proposed a hybrid meta-atom resulting from a combination of a cross and its complement, and suggested that the use  of the Babinet principle may address the above mentioned problems with twisted structures~ \cite{Hannametal2013}. We predicted that this structure may have large, dispersionless optical activity at the transmission resonance, accompanied by very low ellipticity. A numerical study of a similar structure was reported recently for the THz regime~\cite{Zhuetal2013}. Importantly, such structures have $\mathrm {C}_4$ symmetry, which is of lower-order symmetry than commonly studied metamaterial structures created by twisted identical resonators, which have $\mathrm {D}_4$ symmetry~\cite{Deckeretal2009,Lietal2011}.

To further understand this new type of metasurfaces, it is very important to calculate the effective parameters of such structures. Obtaining the material parameters of metamaterial structures is a well established procedure for isotropic, achiral media~\cite{Smithetal2002}. The approach has been extended for the cases of chiral, bianisotropic and inhomogeneous media~\cite{Zhaoetal2010,Lietal2009,Kildishevetal2011,Smithetal2005}. An alternative approach based on the state-transition matrices has also been proposed for isotropic chiral media~\cite{Zarifietal2013}, however none of these methods can be employed for the case of the $\mathrm {C}_4$ symmetry group. The parameters for the structures with $\mathrm {C}_4$ symmetry were retrieved in Ref.~[\onlinecite{Zhangetal2009}] under the assumption that the two bi-anisotropic parameters are related by a frequency-independent constant. This assumption is not valid for general structures, including the one proposed here. This lower-symmetry results in the reflection being dependant on the propagation direction, and is due to the structure being physically different when seen from opposite directions.

\begin{figure}[tb]
	\centering
		\includegraphics[width=\columnwidth]{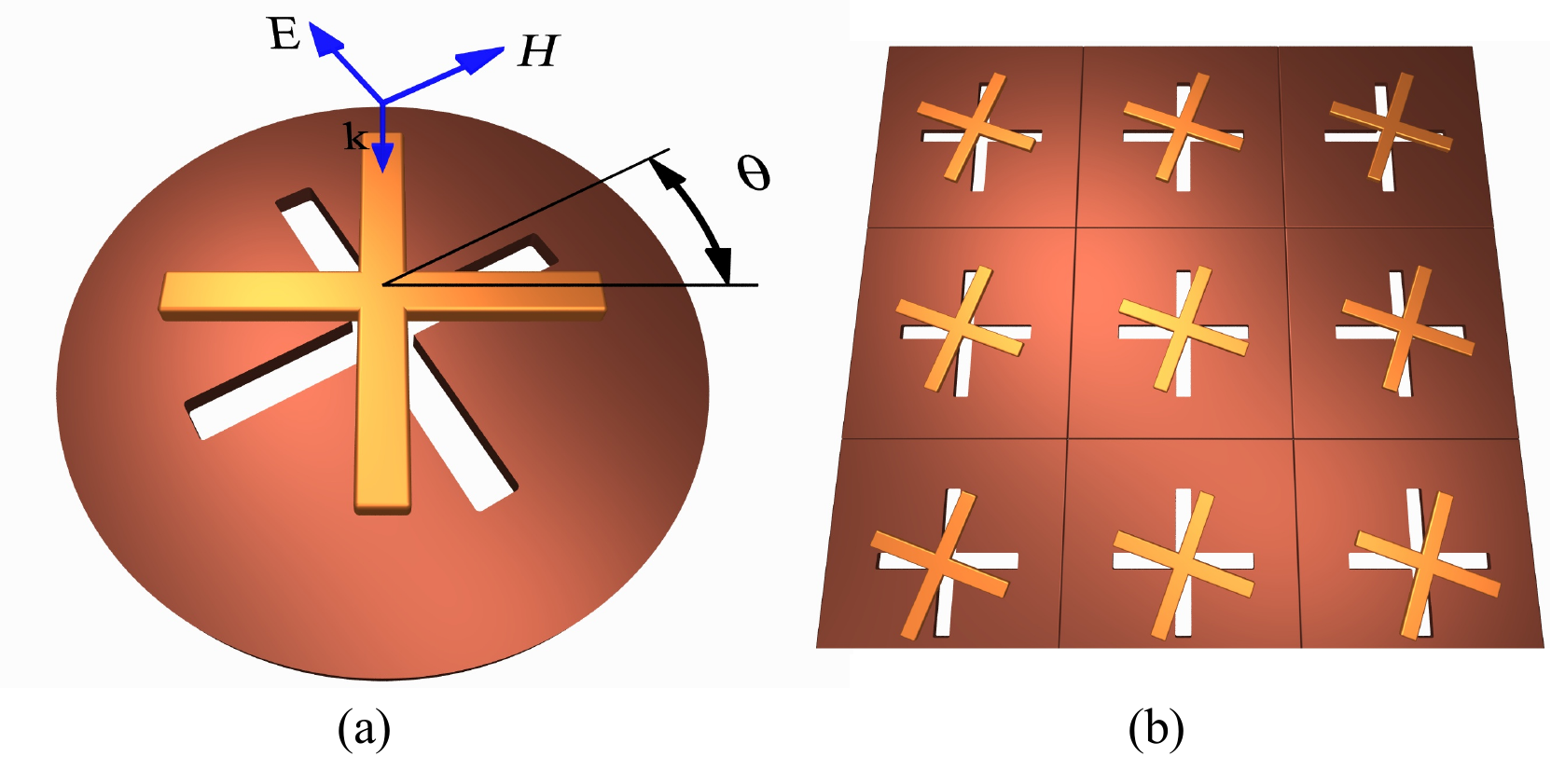}
	\caption{(a) Schematic of the hybrid structure: a cross coupled to its complement, rotated through an angle $\theta$. (b) Our structure in a unit cell configuration.}
	\label{fig:Schem}
\end{figure}

In this paper, we study theoretically and experimentally the properties of metamaterials composed of twisted pairs of cross-shaped meta-atoms and their complements, and develop a retrieval procedure to determine the effective material parameters for the meta-structures with $\mathrm {C}_4$ symmetry. We verify our new theoretical approach by reproducing both numerical and experimental scattering parameters.

The paper is organized as follows. In Sec.~\ref{sec:experiment} we experimentally verify our previous results, finding a good agreement with numerical simulations and confirming our previous findings. We then develop an approach in Sec.~\ref{sec:retrieval} to retrieve the effective parameters for such structures in a unit cell configuration, based on the eigenvalues of the scattering-transfer matrix. Finally, we verify this approach by recalculating the scattering matrix through the substitution of the retrieved material parameters. Section~IV concludes the paper. 


\section{Experimental results}
\label{sec:experiment}

\begin{figure}[tb]
	\centering
		\includegraphics[width=\columnwidth]{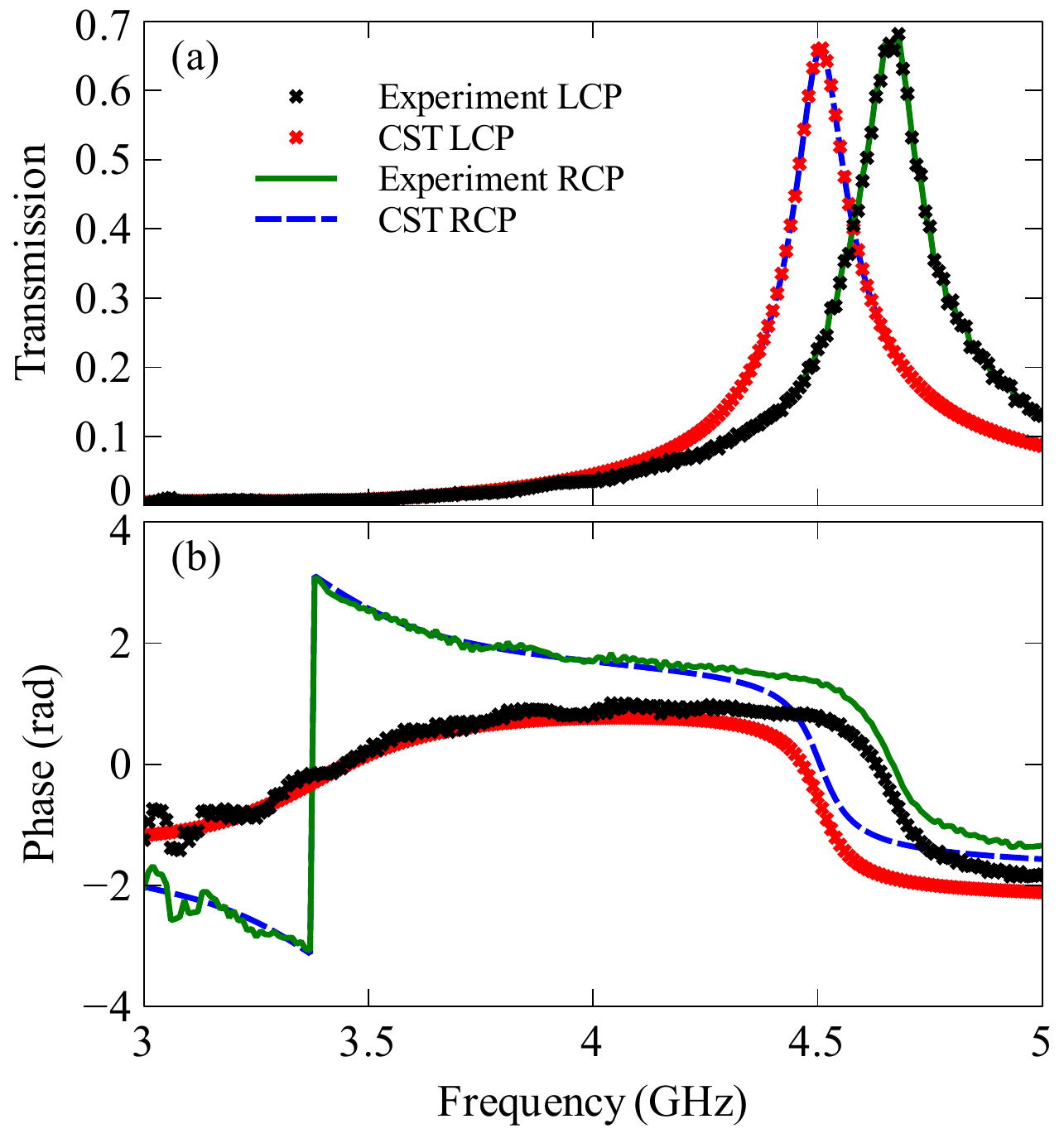}
	\caption{(a) Magnitude and (b) phase of the LCP and RCP waves,  both experimental and numerical, for structure rotated through $20^{\circ}$.}
	\label{fig:trans}
\end{figure}

We choose the cross and its complement to have arms of length $27$mm and width $1.5$mm. They are separated by a Rogers R4350 board $1.5$mm thick, with dielectric constant $3.48$, and twisted through $20^{\circ}$. The metal components are made of copper, $30\mu$m thick. We conduct the experiment inside a circular waveguide, measuring the scattering matrix for both right- and left-handed polarizations.  A schematic of the two elements rotated through an angle $\theta$ is shown in Fig.~\ref{fig:Schem}(a).

Simulations are performed using CST Microwave Studio, using a linearly polarized input wave propagating along the $z$-axis, where the first two polarization-degenerate modes are excited. The first mode is assigned to that with the electric field oriented along the $y$-axis, and the second along the $x$-axis. We simulate the co- and cross-polarized transmission coefficients for both linear polarizations ($S_{xx}$, $S_{yy}$, $S_{xy}$ and $S_{yx}$), and use these to calculate the transmission for the two circularly polarized waves. As our structure has four-fold rotational symmetry, $S_{yy} = S_{xx}$ and $S_{xy} = -S_{yx}$. The magnitudes of the right- and left-handed polarizations are compared with the experimental results in Fig.~\ref{fig:trans}(a). We see that there is little difference between the two polarizations, however the resonances are blue-shifted in the experiment, which is most likely due to imperfect electrical connection between the metallic sample and the waveguide walls. We also plot the phase for both polarizations in Fig.~\ref{fig:trans}(b), and see good agreement apart from the shift in resonance.

\begin{figure}[tb]
	\centering
		\includegraphics[width=\columnwidth]{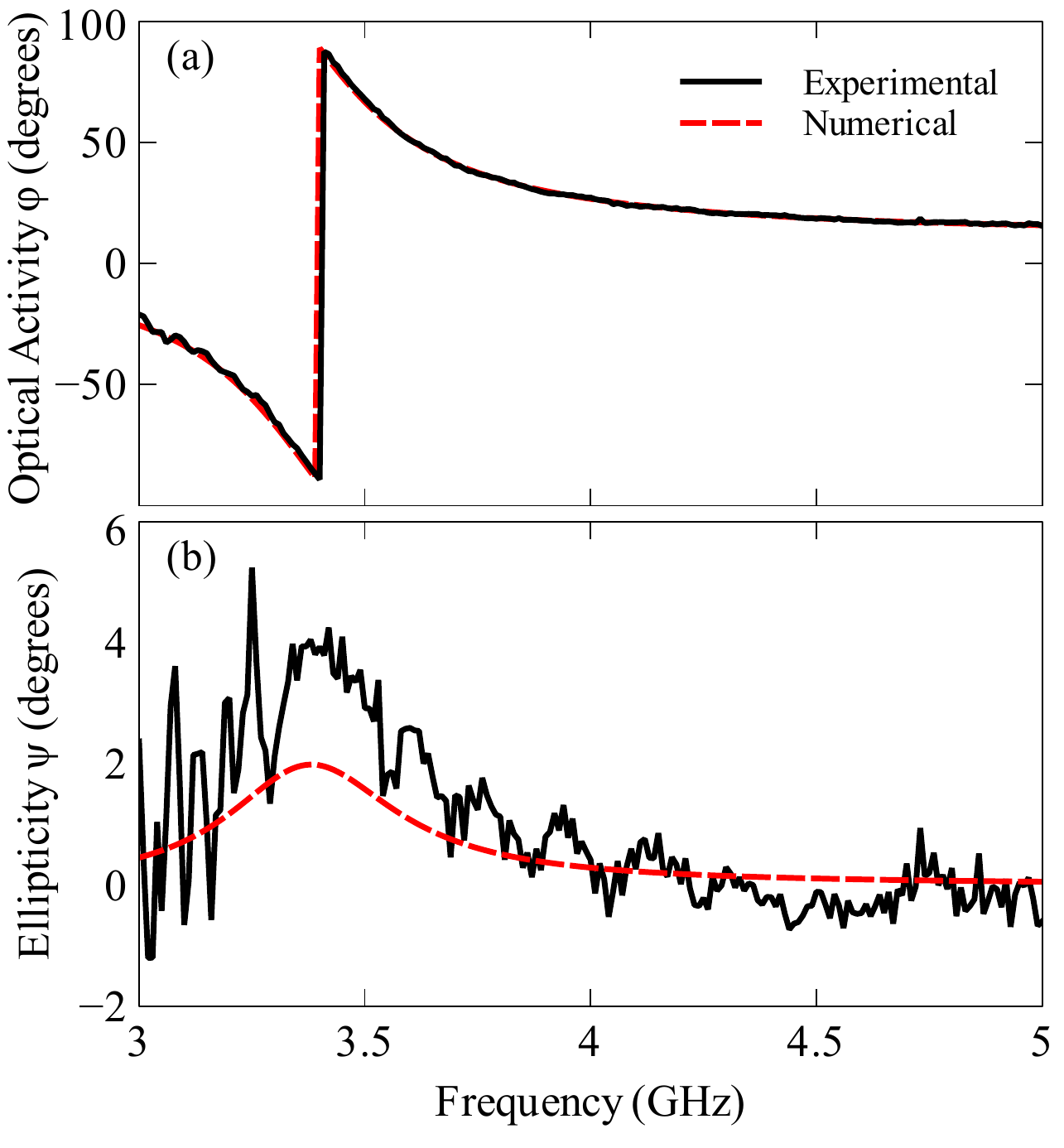}
	\caption{(a) Optical activity for both numerical simulations and experiment, when $\theta = 20^{\circ}$. (b) Experimental and numerical ellipticity, for $\theta = 20^{\circ}$.}
	\label{fig:OA}
\end{figure}

The optical activity is related to the difference in phase between the two polarized waves, while the ellipticity to the difference in transmission magnitudes. We calculate these values using the equations outlined in Ref.~[\onlinecite{Hannametal2013}]. Fig.~\ref{fig:OA}(a) shows the calculated optical activity, comparing the experiment with the numerical simulations. We see that we have good agreement, and see large, flat optical activity over the frequency of transmission. The ellipticity is plotted in Fig.~\ref{fig:OA}(b), noting that the magnitude of the ellipticity is very small, as intended with this design, so the measured values are comparable to the experimental uncertainties. As the ellipticity corresponds to the gradient of the optical activity, it is not surprising that we see very low ellipticity in the region of transmission resonance, accompanying the low dispersion in the optical activity. These results are consistent with our previous findings, where we compared the response of our mixed structure against that of a pair of crosses and a pair of complementary crosses\cite{Hannametal2013}.

\begin{figure}[tb]
	\centering
		\includegraphics[width=\columnwidth]{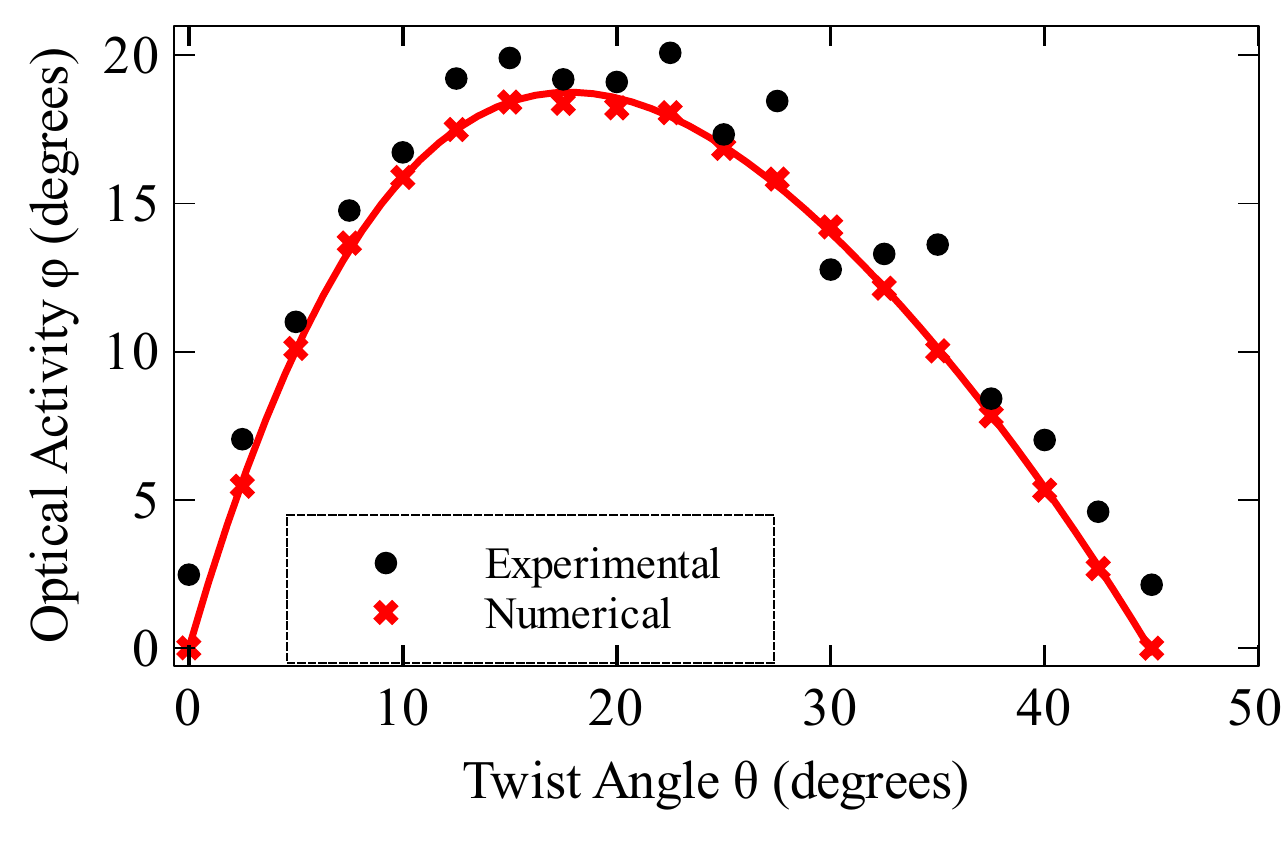}
	\caption{Experimentally measured and numerically calculated optical activity at
 the transmission resonance, as a function of the twisting angle $\theta$. The numerical values are fitted using a $4$th order polynomial.}
	\label{fig:twist}
\end{figure}

Since the system is achiral when $\theta = 0^{\circ}$ or $45^{\circ}$, we expect that by changing $\theta$ we can control the optical activity. We measured the transmission for $\theta = 0^{\circ}$ to $45^{\circ}$, in $2.5^{\circ}$ steps. The resulting optical activity at the transmission resonance is plotted as a function of $\theta$ in Fig.~\ref{fig:twist}, both numerically and experimentally, showing that the optical activity is highly dependent on the twist angle. The small disagreement between numerics and experiment can be explained by imperfections in the fabrication. We also see, from the numerical simulations, that the angle of maximum optical activity is actually about $17.5^{\circ}$, while we would expect it to be at $22.5^{\circ}$ as that is the angle that the system is furthest away from a symmetric configuration. The reason for this discrepancy is the retardation over the gap between the elements, as explained further in Ref.~[\onlinecite{MLiuetal2012a}].

These experimental results verify our previous numerical findings of large, dispersionless optical activity at resonance, and very low ellipticity\cite{Hannametal2013}.

\section{Retrieval of the effective parameters}
\label{sec:retrieval}

To calculate the material parameters we use a unit cell model periodic in the $x$ and $y$ directions for simplification as the waveguide mode is not uniform in the transverse direction, making it equivalent to a non-normal angle of incidence. The cross and its complement are modeled as having arms $28$mm in length and are separated by $1.5$mm. The metal is modeled as PEC. All other parameters remain the same, except that the complementary cross and the boards are now square in shape, to fill up the unit cell, shown in Fig.~\ref{fig:Schem}(b). The system is excited using a plane wave at normal incidence, described using the time convention exp$\left(i\omega t\right)$.

The most general case for our structure, inclusive of all angles, has $\mathrm {C}_4$ symmetry. At normal incidence there is no $z$ component of the macroscopic fields allowing us to model the transverse components using the reduced tensors

\begin{eqnarray}
\bar{\bar{\epsilon}} =
\left(\begin{matrix}
\epsilon & 0\\
0 & \epsilon
\end{matrix}\right), \quad
&
\bar{\bar{\mu}} =
\left(\begin{matrix}
\mu & 0\\
0 & \mu
\end{matrix}\right), \quad
&
\bar{\bar{\kappa}} =
\left(\begin{matrix}
\kappa & \xi\\
-\xi & \kappa
\end{matrix}\right),
\label{eq:kappa}
\end{eqnarray}
where $\epsilon$ is the effective permittivity, $\mu$ the effective permeability, $\kappa$ the chirality, and $\xi$ is a bi-anisotropic parameter which is not present in isotropic chiral media, and is introduced by the lower order of symmetry in our system.
The off-diagonal components of $\bar{\bar{\epsilon}}$ and $\bar{\bar{\mu}}$ are $0$, due to time reversal symmetry~\cite{materials}. The resulting constitutive relations at normal incidence are

\begin{eqnarray}
\left(\begin{matrix}
{\mathbf D}\\
{\mathbf B}
\end{matrix}\right)
=
\left(\begin{matrix}
\epsilon\bar{\bar{I}} & -i/c\left(\kappa\bar{\bar{I}} -\xi\bar{\bar{J}}\right)\\
i/c\left(\kappa\bar{\bar{I}}+\xi\bar{\bar{J}}\right) & \mu\bar{\bar{I}}
\end{matrix}\right)\cdot
\left(\begin{matrix}
{\mathbf E}\\
{\mathbf H}
\end{matrix}\right),
\label{eq:NewCon}
\end{eqnarray}
where $\bar{\bar{J}} = {\mathbf z}_0\times\bar{\bar{I}}$ is the $90^{\circ}$ rotator in the $x -y$ plane. We then have the following parameters to calculate: $\epsilon$, $\mu$, $\kappa$ and $\xi$. The currently established approaches do not cover general structures with this particular symmetry\cite{Zhaoetal2010,Smithetal2005,Zhangetal2009}, so we need to develop a new approach. We have the added complication that due to the meshing in the CST model not preserving $90^{\circ}$ rotational symmetry the eigenstates are not perfectly circularly polarized in the numerical model. To account for this we develop a much more robust method, where we find the scalar parameters of the eigenmodes of the scattering-transfer matrix and use them to assign effective parameters for a medium with circular eigenstates.

\subsection{Eigenmode analysis}
\label{sec:nz}

We start by solving the eigenvalues of the scattering-transfer matrix which are then used to find the refractive index $n$ and the impedance $Z$. The impedance is a tensor, but due to symmetry there are only a few unique values which we will find. When dealing with the tensors, we will denote $\overset{\Rightarrow}{Z}$ as the impedance for waves travelling in the $+z$ direction, and $\overset{\Leftarrow}{Z}$ in the $-z$ direction. To calculate $n$ and $Z$ from the scattering parameters, we make use of the scattering-transfer matrix\cite{scattering}, which can be found from the scattering matrix
\begin{eqnarray}
{\mathbf {T}}_S =
\left[\begin{matrix}
{\mathbf {S_{21}}}^{-1} & & -{\mathbf {S_{21}}}^{-1}{\mathbf {S_{22}}}\\
{\mathbf {S_{11}}}{\mathbf {S_{21}}}^{-1} & & {\mathbf {S_{12}}}-{\mathbf {S_{11}}}{\mathbf {S_{21}}}^{-1}{\mathbf {S_{22}}}
\end{matrix}\right].
\end{eqnarray}
${\mathbf {S_{11}}}$, ${\mathbf {S_{12}}}$, ${\mathbf {S_{21}}}$ and ${\mathbf {S_{22}}}$ are $2\times 2$ arrays including both linear polarizations at each port. We then find the eigenvalues $\lambda_n$ of ${\mathbf {T}}_S$, by using the relation
\begin{equation}
{\mathbf F}(z+d) = {\mathbf T}_S{\mathbf F}(z) = e^{i\alpha d}{\mathbf F}(z),
\label{eq:phase}
\end{equation}
where $\alpha$ is the phase advance across the unit cell of thickness $d$, and ${\mathbf F}$ is defined as
\begin{eqnarray}
{\mathbf F}(z+d) =
\left(\begin{matrix}
b_1\\
b_2\\
a_1\\
a_2
\end{matrix}\right),
\label{eq:eigvec}
\end{eqnarray}
where $a_n$ and $b_n$ are the amplitudes of the waves propagating towards and away from the structure, and can be defined as
\begin{eqnarray}
b_n = (\sqrt{z_0}E_n + H_n/\sqrt{z_0})/2,
\label{eq:bn}\\
a_n = (\sqrt{z_0}E_n - H_n/\sqrt{z_0})/2.
\label{eq:an}
\end{eqnarray}
The value of $n$ refers to the mode being considered and $z_0$ is the impedance of free space. 

\begin{figure}[tb]
	\centering
		\includegraphics[width=\columnwidth]{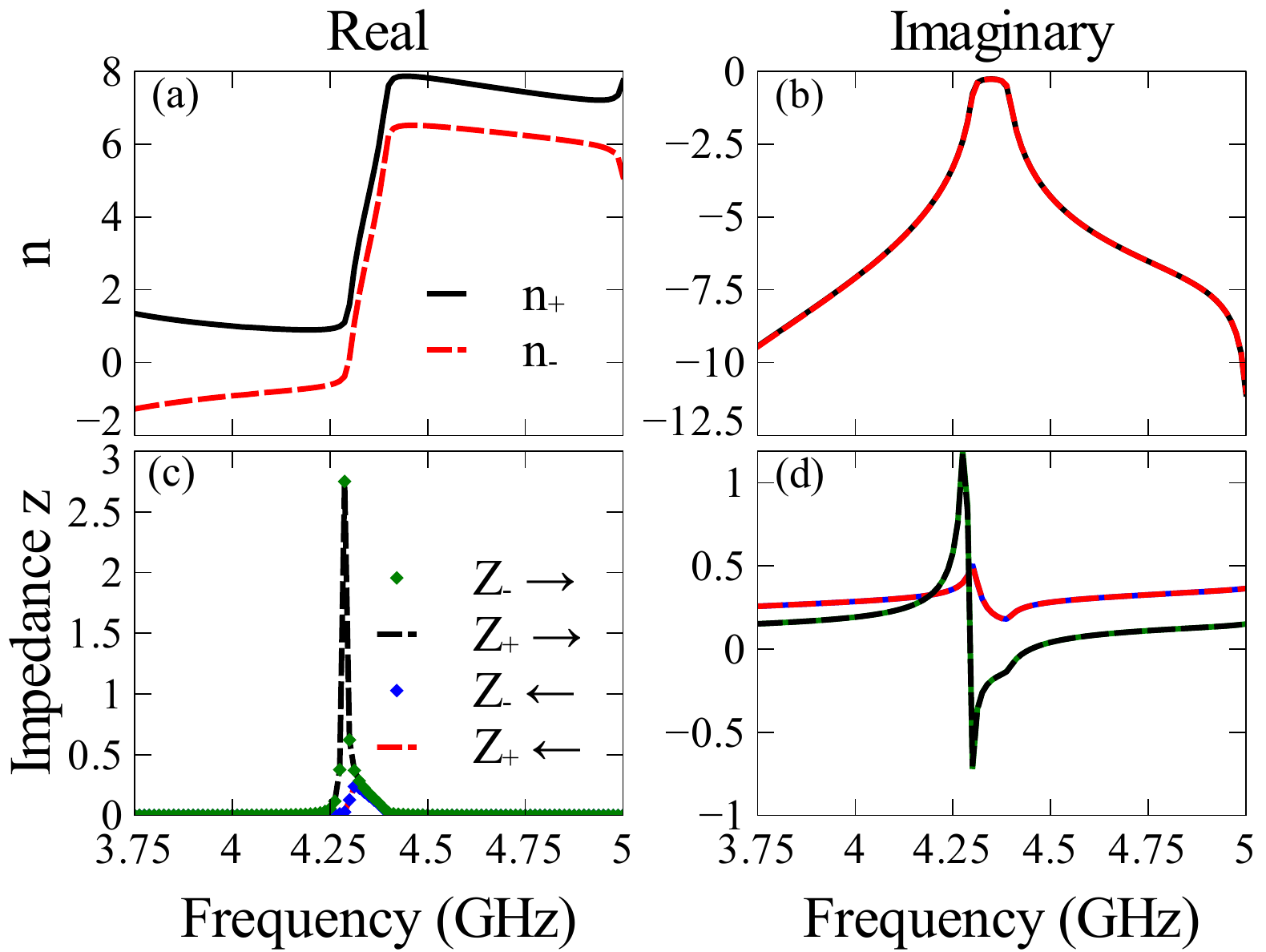}
	\caption{(a) Real and (b) imaginary parts of the retrieved refractive indices for both polarizations. (c) Real and (d) imaginary impedances for both polarizations, for both forward and backward directions. The forwards direction is denoted by $\rightarrow$, and the backwards by $\leftarrow$.}
	\label{fig:nz}
\end{figure}

Reference~[\onlinecite{Smithetal2005}] uses the relation in Equation~(\ref{eq:phase}) with the transmission matrix, however this still holds when using scattering transfer parameters as well.  The four eigenvalues correspond to the forwards and backwards modes of the two polarizations. The refractive indices can then be found as
\begin{equation}
n = \frac{ln(\lambda)}{k_0 d},
\end{equation}
where $d$ is the thickness of the sample (the substrate thickness plus the thickness of both metal resonators), and $k_0$ is the wavenumber in free space. The resulting indices for the forwards direction of the two polarizations are plotted in Fig.~\ref{fig:nz}(a) and (b).

By finding the eigenvectors ${\mathbf F}$ corresponding to these eigenvalues, we can study the fields in the structure. We can determine the eigenstates in our structure by looking at the eigenvectors (not shown). The eigenstates are almost circularly polarized. Equations~(\ref{eq:eigvec}) -~(\ref{eq:an}) can be rearranged to find the ratio of $E/H$ in order to calculate the scalar impedances. For circularly polarized waves the impedances are found as

\begin{eqnarray}
Z^{\pm} = \frac{E^{\pm}}{H^{\pm}} = z_0\left[\frac{b_2+a_2\pm i\left(b_1+a_1\right)}{b_2-a_2\pm i\left(b_1-a_1\right)}\right].
\label{eq:zeig}
\end{eqnarray}
$b_1$, $b_2$, $a_1$ and $a_2$ each have unique values corresponding to each of the eigenvalues, and the values are chosen accordingly.

The resulting impedances are plotted in Fig.~\ref{fig:nz}(c) and (d), for a twist angle of $20^{\circ}$. We see that the impedances are only dependant on the propagation direction. This is expected, as in more symmetric chiral materials, the two circular polarizations have the same impedance\cite{Zhaoetal2010}.

\subsection{Parameter retrieval}
\label{sec:parameters}

Now that we have the scalar index of refraction and impedance for each eigenmode, we can calculate the effective medium parameters. Using equations (8.6 - 8.10) from Ref.~[\onlinecite{materials}] modified for a plane wave at normal incidence, we find the refractive index $n$ of the two circular polarizations in the form

\begin{equation}
n_{\pm} = \sqrt{\epsilon\mu-\xi^2}\pm\kappa.
\label{eq:n}
\end{equation}

We can then find the impedance from equations (8.6), (8.7) and (8.38) from Ref.~[\onlinecite{materials}], by assuming a plane wave in the form exp$(-ink_0d)$ at normal incidence.

\begin{eqnarray}
\overset{\Rightarrow}{Z}_{1,2} = \frac{\eta_0}{\epsilon}\left[\left(ik_0\xi-n_{\mp}\right)\bar{\bar{I}}  + ik_0\kappa\bar{\bar{J}}\right],
\\
\overset{\Leftarrow}{Z}_{1,2} = \frac{-\eta_0}{\epsilon}\left[\left(ik_0\xi+n_{\pm}\right)\bar{\bar{I}}  + ik_0\kappa\bar{\bar{J}}\right].
\end{eqnarray}

We can find the eigenvalues z for the different polarizations and propagation directions, which give us the impedances for the eigenstates in the medium. For $\overset{\Rightarrow}{Z}_{1,2}$ we get
\begin{eqnarray}
z_1 = \frac{\eta_0}{\epsilon}\left(i\xi-n\right),
\label{eq:eigenvalues1}
\end{eqnarray}
and for $\overset{\Leftarrow}{Z}_{1,2}$
\begin{eqnarray}
z_2 = -\frac{\eta_0}{\epsilon}\left(i\xi+n\right),
\label{eq:eigenvalues2}
\end{eqnarray}
where
\begin{equation}
n = \frac{n_+ + n_-}{2}.
\end{equation}
We see that of the four eigenvalues, only two are unique. This supports our earlier argument that the impedance is only dependent on the direction, as shown in Fig.~\ref{fig:nz}(c-d).

Using these eigenvalues, we can rearrange them to find equations for the retrieval of the parameters $\mu$, $\epsilon$, $\kappa$ and $\xi$:
\begin{align}
\epsilon = \frac{2\eta_0 n}{z_1 + z_2}; &
& \xi = \frac{i\epsilon(z_2 - z_1)}{2\eta_0};\\ \nonumber
\mu = \frac{n^2+\xi^2}{\epsilon};
 & & \kappa = \frac{n_{+}-n_{-}}{2}.
\label{eq:parametersFullCase}
\end{align}

\begin{figure}[tb]
	\centering
		\includegraphics[width=\columnwidth]{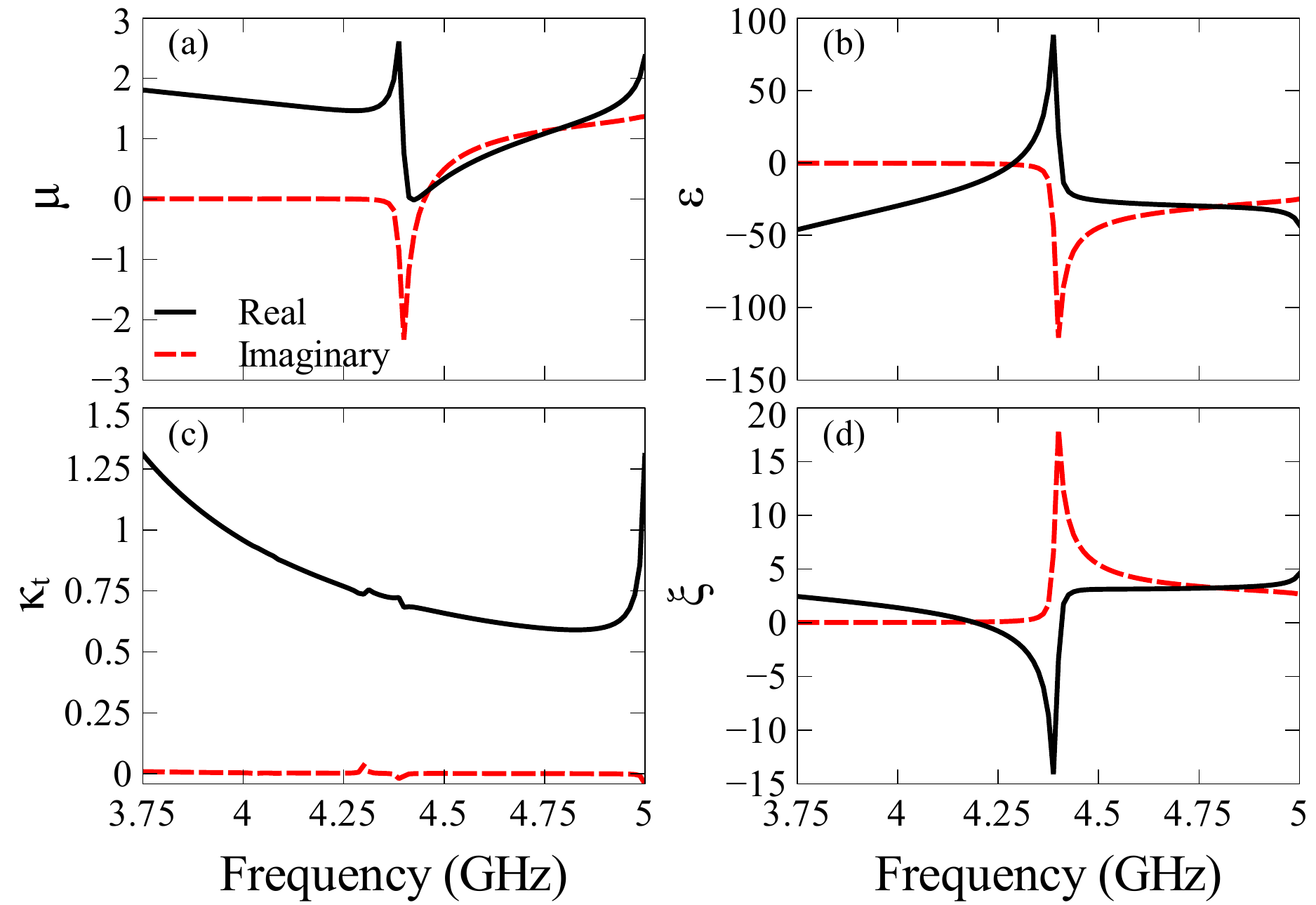}
	\caption{Real and imaginary parts of the retrieved parameters: (a) $\mu$, (b) $\epsilon$, (c) $\kappa_t$ and (d) $\xi$.}
	\label{fig:parameters}
\end{figure}

Both the real and imaginary parts of these retrieved parameters are plotted in Fig.~\ref{fig:parameters}. In Fig.~\ref{fig:parameters}(a) we see that the imaginary part of $\mu$ becomes positive, which violates passivity. However this is a known problem with assigning local parameters to metamaterials\cite{Smithetal2002}, despite which the effective parameters can still yield useful insights. In Fig.~\ref{fig:parameters}(c) we have $\kappa$, the real part of which is directly related to the optical activity, and the imaginary part defines the ellipticity. We see relative flatness in the real part, which is consistent with our earlier findings with the optical activity. We also see that the imaginary part is very low, corresponding to the very low ellipticity reported.

The real and imaginary parts of $\xi$ are plotted in Fig.~\ref{fig:parameters}(d). This reproduces the asymmetry of the structure as shown in the reflection coefficients.

\begin{figure}[tb]
	\centering
		\includegraphics[width=\columnwidth]{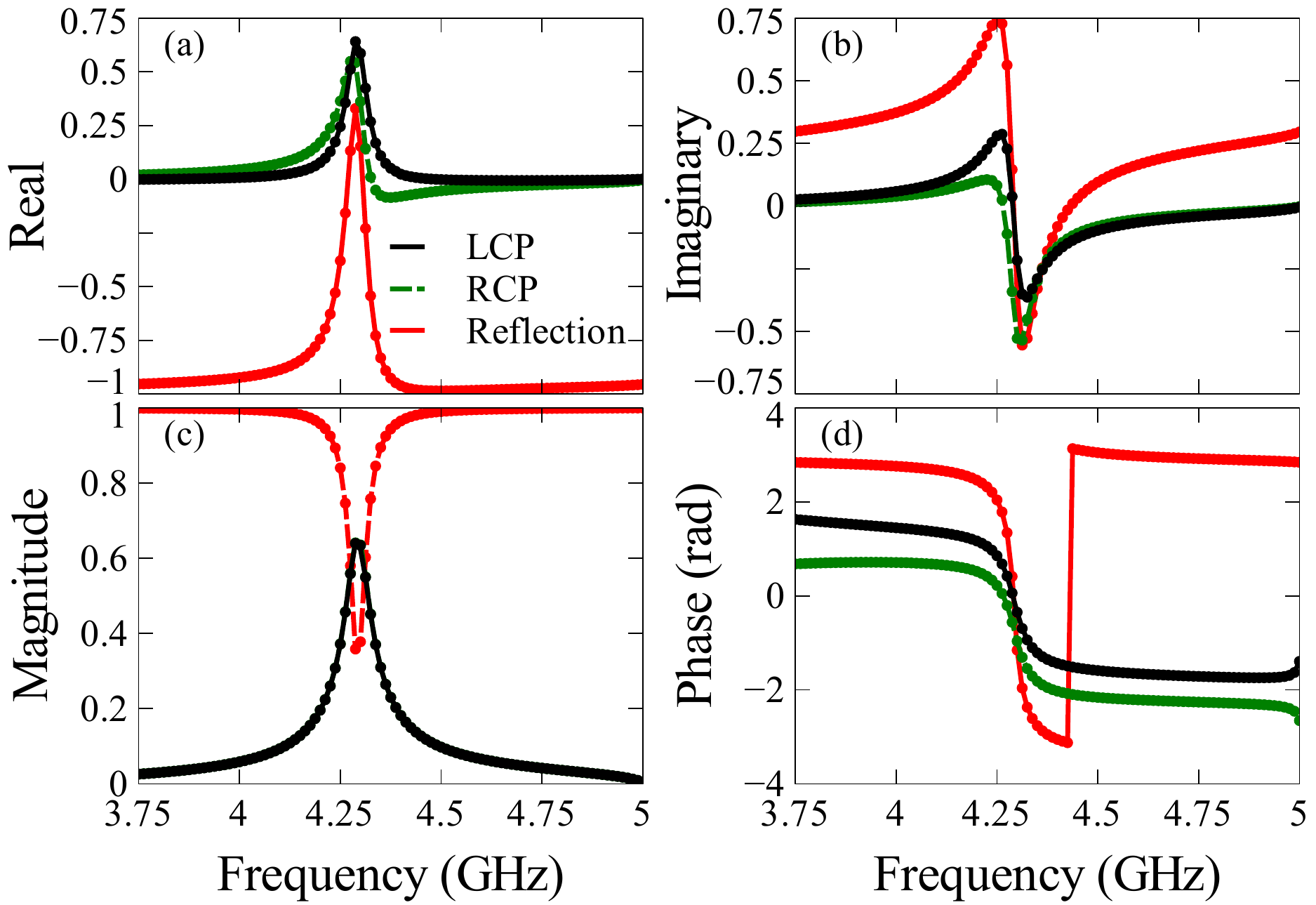}
	\caption{(a) Real, (b) imaginary, (c) magnitude and (d) phase of the scattering parameters calculated by re-substituting the retrieved parameters. The lines are from CST, the markers from the effective parameter model.}
	\label{fig:scat}
\end{figure}

In order to verify the accuracy of this approach, we used our retrieved parameters to recalculate the scattering parameters by re-substitution, using equation (8.39) from Ref.~[\onlinecite{materials}] to calculate the admittance, then equations (8.40-8.46), (8.51-8.52) to calculate the scattering parameters. The results for both polarizations are plotted in Fig.~\ref{fig:scat}, and show near perfect agreement between our original simulations and the recalculations. We can also see the nearly constant difference between the transmission phases in Fig.~\ref{fig:scat}(d), consistent with the flat optical activity. The reflection plotted is that for the forward incidence - to recalculate the opposite direction, the sign of $\xi$ needs to be changed. These calculated scattering parameters confirm the accuracy of our retrieval approach, and also justify us treating the polarizations of the eigenmodes as circular, as this is the assumption made in calculating the parameters.

\section{Conclusions}
\label{sec:conclusion}

We have demonstrated experimentally that the metasurface composed of twisted pairs of meta-atoms with their complement exhibits large, flat optical activity and very low ellipticity. We have studied the response of our structure to a changing twist angle and found the optimal twist angle for maximum optical activity. Because this metasurface has $\mathrm {C}_4$ symmetry, we have developed a novel retrieval method for calculating the effective material parameters which is applicable to structures with $\mathrm {C}_4$ symmetry.  This approach can be easily extended for use in  more general media, potentially including structures inside a waveguide. We have verified the accuracy of this approach by calculating the scattering parameters theoretically and comparing them with results obtained from numerical simulations and experiment.

\section*{Acknowledgements}

This work was supported by the Australian Research Council, the Australian National University, and the Ministry of Education and Science of the Russian Federation.

\end{document}